\documentclass{article}
\usepackage{xcolor,colortbl}
\usepackage{color,soul}
\definecolor{Gray}{gray}{0.85}
\newcolumntype{C}{>{\centering\let\newline\\\arraybackslash\hspace{0pt}}m{0.1\textwidth}}
\newcolumntype{A}{>{\centering\let\newline\\\arraybackslash\hspace{0pt}}m{0.05\textwidth}}
\newcolumntype{V}{>{\centering\let\newline\\\arraybackslash\hspace{0pt}}m{0.4\textwidth}}
\newcolumntype{L}{>{\let\newline\\\arraybackslash\hspace{0pt}}m{0.85\textwidth}}
\newcolumntype{B}{>{\let\newline\\\arraybackslash\hspace{0pt}}m{0.82\textwidth}}
\newcolumntype{M}{>{\let\newline\\\arraybackslash\hspace{0pt}}m{0.1\textwidth}}
\usepackage{float}
\usepackage{comment}
\usepackage{placeins}
\usepackage{makecell}
\usepackage{array}
\usepackage{comment}
\usepackage[nolist]{acronym}

\usepackage{subfig}
\usepackage{amsmath}
\usepackage{enumitem}
\usepackage{longtable}
\usepackage{xcolor}
\usepackage{comment}

\usepackage{tcolorbox}
\usepackage{framed}

\usepackage{url}
\usepackage{enumitem}
\usepackage{placeins}
\usepackage{xcolor,colortbl}
\usepackage{color,soul}
\definecolor{Gray}{gray}{0.85}

\usepackage{float}
\usepackage{comment}
\usepackage{placeins}
\usepackage{makecell}
\usepackage{array}
\usepackage{comment}
\usepackage[nolist]{acronym}

\usepackage{amsmath}
\usepackage{enumitem}
\usepackage{longtable}
\usepackage{xcolor}

\usepackage{arxiv}

\usepackage[utf8]{inputenc} 
\usepackage[T1]{fontenc}    
\usepackage{hyperref}       
\usepackage{url}            
\usepackage{booktabs}       
\usepackage{amsfonts}       
\usepackage{nicefrac}       
\usepackage{microtype}      
\usepackage{lipsum}		    
\usepackage{graphicx}
\usepackage{natbib}
\usepackage{doi}

\usepackage{enumitem}
\usepackage[nolist]{acronym}
\usepackage{todonotes}
\usepackage{pdfpages}

\newcommand{\trtitle}{Who Owns The Robot?: Four Ethical and Socio-technical Questions about Wellbeing Robots in the Real World through Community Engagement}
\title{\trtitle}

\author{
    Minja Axelsson\textsuperscript{\rm 1},
    Jiaee Cheong\textsuperscript{\rm 1,3},
    Rune Nyrup\textsuperscript{\rm 2},
    Hatice Gunes\textsuperscript{\rm 1}\\
    \textsuperscript{\rm 1}University of Cambridge, UK \\
    \textsuperscript{\rm 2}Aarhus University, Denmark \\
    \textsuperscript{\rm 3}Harvard University, USA \\ 
 	\texttt{\{mwa29, jc2208, hg410\}@cam.ac.uk,
rune.nyrup@css.au.dk}
}

\renewcommand{\shorttitle}{Who Owns The Robot?: Four Ethical and Socio-technical Questions about Wellbeing Robots in the Real World through Community Engagement}

\hypersetup{
pdftitle={\trtitle},
pdfsubject={},
pdfauthor={Minja~Axelsson, Nikhil~Churamani, Atahan~Caldir and Hatice~Gunes},
pdfkeywords={Robot Ethics, AI Ethics, Human-Robot Interaction, Community-centred Research, Wellbeing, User-centred Research, Socio-technical Research, Robot Design, AI Policy, Socially Assistive Robotics},
}

\usepackage{bibentry}

\begin{document}

\maketitle

\begin{abstract}
Recent studies indicated that 
robotic coaches can play a crucial role in promoting wellbeing. However, the real-world deployment of wellbeing robots raises numerous ethical and socio-technical questions and concerns. 
To explore these questions, we undertake a community-centered investigation to examine three different communities' perspectives on the ethical questions related to using robotic wellbeing coaches in real-world environments. We frame our work as an anticipatory ethical investigation, which we undertake to better inform the development of robotic technologies with communities' opinions, with the ultimate goal of aligning robot development with public interest. 
In our study, we conducted interviews and workshops with three communities who are under-represented in robotics development: 1) members of the public at a science festival, 2) women computer scientists at a conference, and 3) humanities researchers interested in history and philosophy of science. 
In the workshops, we collected qualitative data by using the Social Robot Co-Design Canvas on Ethics, which participants filled in individually. We used this tool as it is designed to investigate ethical issues of robots with multiple stakeholders. We analysed the collected qualitative data with Thematic Analysis, informed by notes we took during the workshops.
Through our analysis, we identify four themes regarding key ethical and socio-technical questions about the real-world use of wellbeing robots. We group participants' insights and discussions around these broad thematic questions, discuss them in light of state-of-the-art literature, and highlight areas for future investigation. Finally, we provide the four questions as a broad framework that roboticists can and should use during robotic development and deployment, in order to reflect on the ethics and socio-technical dimensions of their robotic applications, and to engage in dialogue with communities of robot users. The four questions are: 1) Is the robot safe and how can we know that?, 2) Who is the robot built for and with?, 3) Who owns the robot and the data?, and 4) Why a robot?.  

\vspace{0.3mm}

\textbf{Keywords:} Robot Ethics, AI Ethics, Human-Robot Interaction, Community-centred Research, Wellbeing, Socio-technical Research, Robot Design, Socially Assistive Robotics 
\end{abstract}

%

\section{Introduction}

Robotic wellbeing coaches have been recently investigated as means to maintain and improve participants' wellbeing through various wellbeing practices (\cite{spitale2022affective}), such as positive psychology (\cite{spitale2023robotic, jeong2020robotic, jeong2023deploying, axelsson2025participant}) and mindfulness (\cite{bodala2020creating, axelsson2023robotic, matheus2025ommie}). Robots for wellbeing are typically socially interactive, embodied, and apply types of Artificial Intelligence (AI) to engage in interactions with users (\cite{mahdi2022survey}). Such robots form part of and are themselves complex socio-technical systems, i.e., systems in which technology and society interact (\cite{Whitworth_Ahmad_2014}). Researchers' reasoning behind creating such wellbeing robots and using them in the real-world is creating accessible, low-barrier mental wellbeing support (\cite{scoglio2019use, axelsson2021participatory}). However, as robotic technologies are currently rarely used outside of the lab, research on the ethical implications of their real-world use is limited.

%


In concurrence, there has been a proliferation of research focusing on \textit{human-centred ethical AI} (\cite{van2022human,loi2019co,capel2023human}).
Existing works have relied on the definition of ethical AI as ``AI that seeks accountability regarding fundamental human values and rights, and advocates for more transparent design of AI'' (\cite{capel2023human}). 
This includes work in the robotics field, in which various human-centred design processes have engaged prospective users and other stakeholders in examining the ethics of robotic applications (\cite{axelsson2022robots, ostrowski2022ethics, ostrowski2023we}). However, to date, there is still a lack of research conducted on \textit{what} and \textit{how} robots and the related real-world ethical considerations are thought of by under-represented communities. Particularly, members of the public, women computer scientists, and researchers from the humanities who have had limited involvement in the ethical evaluation of robotic technologies. This is due to various structural barriers and biases (\cite{zuger2023ai, hall2023systematic, chun2023crisis}), in computer science, AI and robotics in particular.

%


To further understand these communities' perspectives and to improve inclusivity in the evaluation of ethics,
this work presents three community-based workshops, where participants $(n = 22)$ reflected on the ethical issues related to robotic wellbeing coaching. In the workshops, participants were introduced to robotic wellbeing coaches (either through a live demo or video recordings), and used the Social Robot Co-Design Ethics canvas (\cite{axelsson2021social}) to reflect on ethical issues, filling their own canvases and also engaging in discussions related to the topic with the researcher(s) and each other. This work presents participants' reflections as detailed on the canvases, groups these discussions into four broad ethical and socio-technical questions generated through Thematic Analysis, discusses these thematic questions' relationship to existing literature, and identifies future research directions. Finally, the questions are presented as a tool for roboticists to interrogate and discuss the ethics of robotic applications during robot design and deployment, in conversation with robot users.



\section{Background and Related Work}

Prospective users have highlighted advantages of using robots for wellbeing, such as lack of judgement and accessibility in comparison to human coaches (\cite{axelsson2021participatory, axelsson2022robots}), and professional coaches have highlighted advantages such as reliability, consistency and uniformity in comparison with human coaches (\cite{axelsson2022robots}). In comparison with non-embodied AI (i.e., mobile apps and computer interfaces), prospective users have highlighted the physical presence of a robot as an advantage. This advantage is supported by \citet{sayis2024technology}, who found that in comparison with a voice assistant (i.e., a smart speaker), a wellbeing robot elicited more self-disclosure from participants during a wellbeing exercise, and generated positive changes in mood, whereas the voice assistant did not. 

While these investigations provide a rationale for the potential use of embodied robots to promote wellbeing as an alternative to existing solutions, the investigation of ethical dimensions of robotic wellbeing coaches are still limited. 
\citet{axelsson2022robots} detailed design and ethical recommendations for robotic wellbeing coaches, 
based on qualitative results obtained from three user-centred studies. 
%
%
\citet{spitale2024appropriateness} further investigated the appropriateness of LLM-generated language for a robotic coach via a workshop held with study participants, identifying issues related to ethics and bias in LLMs. However, community engagement about ethical and societal questions if these robots were to be deployed 
is limited.

Moreover, we find that an \textit{anticipatory approach} towards robotic wellbeing coaching ethics is lacking within the field.
Rather than reacting to harms after they occur, anticipatory ethics call for proactive reflection emphasising the identification of potential risks, power imbalances, and unintended consequences early in the design and development process (\cite{barnett2022crowdsourcing,brey2017ethics}). 
Within the scope of our work, this entails deliberating on the future implications (\cite{barnett2022crowdsourcing,rakova2021responsible}) of deploying robotic wellbeing coaches within sensitive use-cases such as mental health and emotional support. 
Given that research on anticipatory governance (\cite{hua2023effective}) and socio-technical harms (\cite{shelby2023sociotechnical}) have demonstrated the effectiveness of such an approach, we put forth that researchers within the scope of wellbeing robots, can and should start investigating the utility of anticipatory ethics within their research. This requires critical examination of who the technology benefits, who may be marginalised, and how the technology aligns with users’ values, cultural contexts, and emotional needs. 
On a deeper level, it involves asking what different---and particularly under-represented---communities perceive about the ethical usage of social robots for wellbeing, concerns that they may have, or subtly shift social expectations about human connection, responsibility, and trust. 

Our work extends these previous works and raises new questions about the ethics of robotic mental wellbeing coaches if they are to be deployed in the real-world as products. \citet{vsabanovic2023robots} previously detailed ``10 defining questions'' to help roboticists identify the strengths and limitations of their robot implementations, in relation to ``good'' robotics.
We take inspiration from that work, and identify the questions raised in this paper through community engagement and workshops, grounded on participants' reflections, to raise thoughts about what users and community members that come into contact with robotic products want to know about the operation of those robots. We define community in a loose sense, in that they have shared interests, but may not necessarily share geographical locality (\cite{bradshaw2008post}).
We view our work to be a contribution toward the growing practice of \textit{critical robotics}, in which ethical and societal ``challenges and dilemmas'' that arise within Human-Robot Interaction (HRI) are critically examined (\cite{serholt2022introduction}). In this frame of reference, we discuss HRI and specifically robots for wellbeing here as complex, socio-technical systems, in which social and technical elements of a complex system are interconnected (\cite{socio-technical}), and can be analysed jointly. 
%


\section{Methodology}

We conducted three community-based workshops on the ethics of robots for wellbeing. Participants $(P_{total} = 22$) were recruited at three different events held at the University of Cambridge. The protocols for each workshop differed slightly, to accommodate the particulars of each event. In each workshop, participants filled in one canvas, the Social Robot Co-Design Canvas on Ethics (\cite{axelsson2021social,Axelsson2020social_canvases}), to elicit their reflections. The canvas addresses six ethical issues explicitly: physical safety, data security, transparency, equality across users, emotional consideration, and behaviour enforcement. The protocol was: 
\begin{itemize}
    \item \textbf{$G_1$:} Members of the \textbf{public at a science festival} at the university $(G_1 = 6)$. First, participants interacted one-on-one with a robot, and could choose either a child-like QTrobot\footnote{https://luxai.com/} (90 cm tall, with heads, a torso, full arms with shoulders, elbows and hands, and static legs, standing on a tabletop) or a toy-like Misty II\footnote{https://www.mistyrobotics.com/misty-ii} (36 cm tall, with head, torso, stubby arms, and tread-like wheels) robot, which did a brief positive psychology exercise. 
    Participants then filled in the canvas, and engaged in a semi-structured interview (approx. 10 minutes) with a researcher.

    \textbf{Demographics:} We did not collect demographics to preserve the privacy of the public attending the science festival, as advised by the departmental Ethics Committee. 
    
    \item \textbf{$G_2$:} Attendees of a \textbf{women in computer science conference} $(G_2 = 12)$, mainly targeted to early career researchers, held at the university. Participants were shown a video of a person engaging in a positive psychology practice with a QTrobot (similar to the interaction demonstrated in G1). Then, they filled the canvas, and engaged in a group discussion (approx. 40 minutes). 
    
    \textbf{Demographics:} 2 participants did not disclose any demographics information, and some omitted some information. The disclosed demographics: 10 participants were female, 5 were aged 18--25, 5 were aged 26--35, and 9 of them reported a computer science background. They had the following nationalities: Chinese, Indian, Romanian, Brazilian, Singaporean, Arab, and British. Six had undergraduates, one had a graduate, and three had PhD degrees. Six had little to no experience with social robots, 2 had some experience, and 2 had frequent work on social robots. 2 had conversational level English, six fluent level, and 2 were native speakers. 
    
    \item \textbf{$G_3$:}  Academics who were part of a \textbf{special interest group in the history and philosophy of science} $(G_3 = 4)$ at the university. Participants were shown the same video as in G2. They then filled their canvases, and engaged in a group discussion (approx. 40 minutes). 
    
    \textbf{Demographics:} Everyone disclosed their demographics. 2 were female and 2 were non-binary, three were aged 26-35 and one was aged 56-65. 2 were in the philosophy field, one in medical sociology, and one in anthropology. They had the following nationalities: Danish, British, Canadian, and UK/USA. 2 had graduates and 2 had PhD degrees. All had little to no experience with social robots. 2 were fluent level and 2 native speakers in English.
\end{itemize}
We chose these spaces and events as they allowed us to engage with communities which are interested but typically under-represented in the development of robotics. 
In order to distil findings, we conducted an inductive qualitative analysis of what the participants wrote on the canvases. We developed our analysis iteratively, broadly following the Thematic Analysis process (\cite{clarke2017thematic}): 1) familiarisation with data, 2) creating initial codes, 3) searching for themes, 4) reviewing the themes, 5) defining and naming the themes, and 6) creating a report. We refined our themes over multiple passes of the data, and in conversation between two of us, to inform the analysis with both our perspectives. We consider our sample size $(n = 22)$ to be sufficient for the purposes of Thematic Analysis, as we observe data saturation (\cite{guest2006many}), and have sufficiently answered our research question (\cite{marshall1996sampling}) about ethical and socio-technical questions.

Both researchers performing the analysis wrote notes during the discussions, which we referred to 
and informed our theme identification. We did this in order to capture participants' thoughts which they did not include on the canvas, and broader conversational themes. We chose not to record the semi-structured interviews or workshops, in order to protect participants' privacy and let them freely engage in the community-based events without feeling observed. 

\begin{figure*}[ht]
    \centering
    \includegraphics[width=0.8\textwidth]{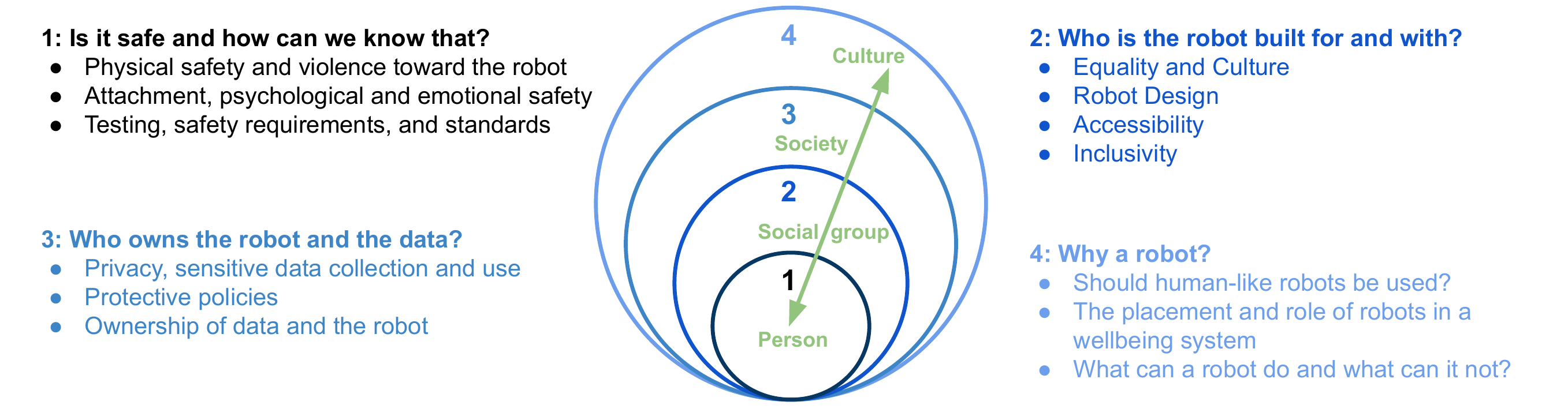}
    {\caption{Results from our three \textit{community-centred anticipatory ethics approach} engagement sessions.
    These layers illustrate how interactions with wellbeing robots are nested within multi-layered systems, from the intimate personal layer to the broader institutional and societal layer. The layers do not have strict boundaries, and are presented here to structure our discussion.
    }
    \vspace{-5mm}
    \label{fig:AIES_2025_results}}
\end{figure*}

We organise our questions into four themes which we present as ethical and socio-technical questions about the real-world use of wellbeing robots. We chose to group our findings specifically as questions, in order to open up conversations related to these themes, and to emphasise the openness of these questions, the lack of straightforward solutions, and the ongoing need to address these questions within the fields of robotics and AI ethics. We have visualised our questions in an expanding circle from the person, to social groups, to broader society and culture, in Figure \ref{fig:AIES_2025_results}. The circles expand from more personal and specific questions, to the more general and abstract. We present this visualisation to aid in structuring and interpreting our findings. However, we do not claim that the themes and sub-questions discussed in relation to, e.g., our first identified ethical question relate only to the innermost layer (i.e., the person), and so forth. This visualisation provides only a starting point for thinking about how the questions raised in this work relate to the individual, citizen, or ``user''; their friends, family, workplace, and other relationships and social groups; local, international and global society; and  local and broader interconnected cultures.  We intend our findings to aid in fostering critical and reflective robot design and development processes, and for robot designers and developers to keep these questions in mind when developing robots, discussing them with users and (under-represented) communities throughout.


\section{The Identified Ethical Questions}

Here, we present the four identified real-world ethical and socio-technical questions, participants' insights and discussions related to each, and discuss them in light of state-of-the-art literature. We highlight areas for future work and make suggestions for how roboticists can begin addressing and reflecting on these questions. 

%

\subsection{Is It Safe and How Can We Know That?}
\label{sec:theme2}

Participants brought up potential safety issues with wellbeing robots, regarding physical and mainly psychological and emotional safety. They discussed e.g., the robot posing a safety risk to vulnerable populations such as children, by them damaging the robot and getting hurt ($G_2P_{01}$, $G_3P_{01}$), or that they may form inappropriate attachments with the robot ($G_1P_{06}$, $G_3P_{02}$). Participants discussed how they might know that a robot is safe, in the form of safety testing and safety requirements. These topics relate mainly to the layer of the ``person'' (see Fig. \ref{fig:AIES_2025_results}).

\subsubsection{Physical Safety and Violence Toward the Robot}

Some participants remarked that the robot's physical form could pose a safety risk, e.g. by being close to the user ($G_2P_{11}$). This was pointed out especially in the case of children, if they were to damage the robot. $G_2P_{01}$ mentioned that the robot could interact ``incorrectly'' with its environment and ``fall off a table or something'', and that ``lots of movements'' could hurt the user ($G_3P_{01}$). $G_1P_{05}$ also mentioned ``getting attacked by the robot'' as a potential risk, potentially indicating an influence from media narratives about robots.

Interestingly, $G_1P_{03}$ mentioned the robot being a safety risk if a person had a ``temper'' and ``lashed out at the robot''. $G_2P_{01}$ also mentioned that ``a user that gets upset may try to damage the robot''. It was not clear from conversations whether participants perceived damaging the robot as a moral wrong in itself, or whether they were primarily concerned about potential harm to the user themselves. Research has found that robots' perceived intelligence seems to influence how willing people are to damage it (\cite{bartneck2008exploring}), and that discriminatory behaviour such as sexist abuse may extend to voice agents (\cite{coalition2019d}) and robots (\cite{winkle2022norm})---thus potentially perpetuating it toward women. Research is needed on whether violence toward robots could perpetuate (gendered or sexualised) violence toward other people, and whether wellbeing robots that inhabit a ``service" role are especially prone to this. 

\subsubsection{Attachment, Psychological and Emotional Safety}

Participants mainly brought up concerns related to emotional and psychological safety (e.g., in relation to attachment) when discussing robots for wellbeing, as opposed to concerns for physical safety. They emphasised that the personal, sensitive and vulnerable nature of discussing wellbeing made these safety concerns important. $G_1P_{05}$ mentioned that ``affection won't be returned'', if a user were to form a bond toward a robot. $G_3P_{01}$ noted that specifically in the context of wellbeing support, people ``would want the robot to recognise them or reciprocate the potential attachment'', and describing a feeling of disappointment if a person had to re-explain who they were to the robot ``that has supported you''. 
$G_1P_{06}$ also mentioned potential risks of emotional attachments by ``putting too much trust in the robot, giving away too much personal information''.
This concern is supported by research: in a study examining a robot in a therapeutic context, participants disclosed more about their personal life when they were feeling stressed and lonely, or had lower mood (\cite{laban2023opening}). 
Such willingness to disclose to robots when in a vulnerable state could potentially be exploited 
for nefarious purposes, e.g., in the case of robot hacking (\cite{winfield2017case}). This is an open question about what kind of information a wellbeing robot should ask for, and what it should not.

Participants also mentioned children being particularly vulnerable users of wellbeing robots. $G_1P_{06}$ mentioned a potential ``critical attachment period in young children'', and that such robots ``should not be used as substitutes for parents away for long period''. $G_3P_{02}$ also questioned whether forming a relationship with a robot could ``warp the development of the child's attachment to humans''. Developers of wellbeing robots for children do not generally aim to substitute parental roles, but rather to support them. As such, their goals and views tend to align with the participants'. As reviewed by \citet{kabacinska2021socially}, robots generally served a supportive role, such as a huggable teddybear or a distraction during a medical procedure. However, it is worth paying attention to how such robots continue to be deployed, and roboticists should advocate for their appropriate positioning as aids and tools of support, rather than any kind of substitute. 

\subsubsection{Testing, Safety Requirements, and Standards}

Participants also questioned how they might know that a wellbeing robot is safe to use. Participants advocated for extensive testing on a cross-section of diverse people to mitigate bias ($G_1P_{04}$, $G_3P_{01}$), iterative design and quality control ($G_1P_{05}$), putting safety requirements into the robot system design ($G_2P_{12}$), and equipping robots with safety measures to avoid and mitigate inappropriate behaviours ($G_2P_{01}$).

While various commercial auditing services and frameworks exist for non-embodied AI (\cite{mokander2023auditing, li2024making}), such approaches for embodied AI (i.e., robots) are still limited. \citet{winfield2023relationship} describe the state of benchmarking, standards, and certification in robotics and AI. They point out the IEEE P700X series as emerging standards for the ethics of autonomous systems, including, e.g., IEEE 7000-2021 Model Process for Addressing Ethical Concerns during System Design and IEEE 7001-2021 Transparency of Autonomous Systems. 
There exists also an ISO standard for personal care robots (ISO 13482:2014). 
However, this standard does not appear to address psychological and emotional safety, and does not apply to robots that are medical devices. 
Additionally, \citet{winfield2023relationship} note that there is a gap in the regulatory landscape with regards to robots used in private homes. 

\begin{framed}
\noindent\textbf{Key Takeaway 1:}
Participants expressed concerns about the physical and emotional safety of wellbeing robots, particularly for vulnerable users like children. Key risks include injury, inappropriate attachment, and misuse of personal data. They emphasised the need for rigorous testing, ethical design, and stronger standards.
\end{framed}


\subsection{Who Is the Robot Built For and With?}
\label{sec:theme3}

The communities 
also posed questions about who the robot is built for and with. These questions are related to factors such as cultural context, personal experience, and the specific design and application of the robots. Participants raised questions about who has been consulted and involved in the design process ($G_1P_{03}$, $G_2P_{03}$), and how well the robot will serve different demographics ($G_3P_{01}$, $G_2P_{03}$). These topics relate mainly to the layer of the ``social group'' (see Fig. \ref{fig:AIES_2025_results}).

%

%


\subsubsection{Equality and Culture}

Participants demonstrated awareness of the current ethical challenges within the AI field with regards to equality and culture.
One example is that of \textbf{AI bias} in HRI.
Recent research has demonstrated that many robots and AI systems learn from biased datasets, which can result in discriminatory algorithmic outputs or behaviours (\cite{hitron2022ai,cheong2021hitchhiker}). 
In their germinal work, \citet{buolamwini2018gender} found that commercial computer vision systems demonstrated intersectional bias in gender classification, i.e., outcomes were least accurate for darker-skinned females. Specifically in the context of robots for wellbeing, \citet{cheong2024small} found that machine learning (ML) bias is present within robot wellbeing coaching datasets collected in-the-wild.
%
A hypothesis for this is that ML algorithms are not trained to account for the inherent difference in depression expression across gender and culture (\cite{cheong2023towards,u-fair_ml4h_2024,cheong2024fairrefuse,cheong2023causal}), thus causing the ML-powered systems to produce biased outcomes when deployed within real-life robot wellbeing coaching use cases.
%
%
Participants proposed approaches to combat bias-related issues. $G_2P_{03}$ for instance proposed we should “feed robots with diverse data collected from different cultures” and to “provide rich appropriate data" to the robot database or training data in order to allow the robots to learn what kind of behaviours are considered acceptable,
%
which are aligned with recommendations from existing works (\cite{cameron2024multimodal,cheong2024small,cheong_acii}).

Participants also highlighted equality concerns over the \textbf{homogenisation of different cultural} groups by treating ``users as white western users and ignore other types of users'' as highlighted by $G_3P_{03}$.  
%
%
Existing literature in HRI have focused on algorithmic fairness (\cite{cheong2024small}), resource distribution (\cite{ostrowski2022ethics}), or robot appearance (\cite{lachemaier2024towards,ogunyale2018does}), but ignores the quality of interaction or appropriateness of robot response across different cultures or identities.
Participants have highlighted how this is a pressing concern. 
For instance, $G_3P_{01}$ emphasised how ``sociocultural differences are very important in the kind of support needed"
and $G_2P_{03}$ was concerned that the robot will ``provide inappropriate advice/comments to some users from different culture''.
Within existing literature, there is evidence that racial
and ethnic minorities tend to receive lower quality health and mental healthcare than non-minorities (\cite{egede2006race,hall2021reducing}) and that other individuals, such as informal caregivers (\cite{kim2024opportunities}) tend to be neglected.
%
%
%
%
%
%
%
This is both an epistemic injustice and a procedural fairness issue, where the system fails to involve diverse user perspectives in its development (\cite{prabhakaran2022cultural}).
The deployment of robots in healthcare and wellbeing must be intersectional and responsive to diverse cultures, needs, roles and identities (\cite{kim2024opportunities,soubutts2024challenges,xie2024empathetic}), and future research with under-represented groups is needed.



\subsubsection{Robot Design}

%
Given the anthropomorphic qualities of social robots, another prominent theme participants have picked up on is the bias present within robot design.
For instance, 
$G_1P_{03}$ questioned ``Why are the robots always white? Where are their eyes being stylised from? Is it based on a white or western norms?''
\citet{haring2018ffab} outlined the implications that a robot's design has on a person's bias to interact socially with a robot.
Beyond the ``whiteness of AI", participant 
$G_3P_{02}$ also questioned “is the robot’s voice gendered? Racialized?''
\citet{hitron2022ai} investigated the effects of a gender-biased robot and its effect on humans' implicit gender stereotypes.
This is aligned with the growing body research that highlights the ``Westernness'' of AI (\cite{howell2025weird,cave2020whiteness}) and the ``whiteness'' of robots (\cite{addison2019robots,strait2018robots,cave2020whiteness,sparrow2019robots}).
%
%

%
Several studies have emphasised 
the necessity to ensure that robots do not reinforce or amplify social inequalities (\cite{ostrowski2023we,zhu2024robots}). 
%
%
These studies suggest that the models’ disparagement of certain groups is not only a reflection of societal biases but also a perpetuation of harmful representations and erasure (\cite{dennler2024designing,skewes2019social}), 
which may lead to a vicious cycle of societal bias amplification (\cite{otegui2025gender, chu2022digital, nyrup2023feminist}). 
%

%

%
%


\subsubsection{Accessibility}
%
Two prominent \textit{accessibility} themes were highlighted. The first is that of \textbf{physical} accessibility. 
%
$G_1P_{03}$ brought up concerns for those with ``hearing issues" and emphasised the need for ``different robotic forms for those with different neurodiversities". 
Existing research also highlighted how robots should accommodate people with disabilities, including those with mobility, vision, or cognitive impairments (\cite{qbilat2021proposal,al2022accessibility}).
Perhaps embedding multilingual capabilities and sign language integration can improve robot inclusivity (\cite{akalin2014non,hei2024bilingual,li2023multimodal,axelsson2019participatory}).

Another accessibility theme mentioned is that of \textbf{affordability}, economic and social availability. 
$G_2P_{01}$ summed this up succinctly by highlighting that “if the robot should appeal to all groups, its dialogue should be relevant for all people otherwise, it can only talk about yoga and green juice. Price affects who it helps”.
%
Themes of inequity due to cost has also been frequently highlighted within existing research (\cite{almuaythir2024robotics}).
It is important that robotic coaches are cost-effective and widely available in order to ensure that underprivileged communities also benefit (\cite{johnson2020task,velor2020low}). Approaches to this could be sharing robots in communities, and building low-cost open source robots (e.g., Blossom, presented by \cite{suguitan2019blossom}). 
%


\subsubsection{Inclusivity}

Another primary theme that emerged 
is that of inclusivity. 
$G_3P_{02}$ pointed out that ``a user that is already experiencing ``bullying” in everyday life can also experience it in the interaction with the robot.
\citet{nomura2020people} found that while people with social anxiety experience less anticipatory anxiety and tension when interacting with a social robot rather than a person, they do still experience those things. Additionally, socially anxious people tend to perceive others' reactions to them more negatively than people with low social anxiety (\cite{pozo1991social}). As robots are prone to errors such as misunderstanding, interrupting and not responding (\cite{spitale2023longitudinal}), this could be particularly disruptive to an anxious person. Research on reducing (\cite{bilac2017gaze}) and repairing (\cite{axelsson2024oh}) these errors could aid in this.

Another challenge that we have picked up on 
 is the inclusivity challenge posed by engineering-centered perspectives within the field of robotics.
Robots designed for mental health support 
may face inclusivity challenges due to a focus on technical and performance-driven priorities, which can overshadow human-centered and ethical considerations (\cite{zhu2024robots}).
%
%
For instance, 
$G_1P_{02}$, who ``admits'' to being ``an engineer" with a very ``practical/ engineering approach towards everything'', believes that ``aesthetics/ cuteness are all just background concerns which which shouldn’t be a primary thing when designing a robotic wellbeing coach.''
This contrasts with some of the other non-engineers (e.g., $G_1P_{01}$, $G_1P_{03}$) who commented that the robot's aesthetics or ``cute factor" helped them to emotionally better relate to the robotic coach.
This could pose a problem. Engineers who prioritise functionality and efficiency over human experiences, emotions, and cultural sensitivities may end up designing or developing robots that are cold, mechanical, or unrelatable to users.
%

%
%
Robot design and development is often dominated by engineering-centred perspectives (\cite{moniz2016robots,zhu2024robots,faulkner2015nuts}), which can lead to it becoming isolated from the societal context in which robots are deployed (\cite{zhu2024robots,faulkner2015nuts}).
This calls into question whether the \textit{design} or the \textit{deployment} of robotic coaches are sufficiently inclusive. 
%
Human-centred and design justice approaches may be highly suited to address this 
(\cite{syal2025design,crivellaros2025co}).
%
In a 10-year survey on affective robots for wellbeing, \citet{spitale2024past} found that user-centred approaches are increasingly being applied to wellbeing robot design. 
The authors called for collaborative design approaches, and the inclusion of multiple stakeholder groups in the design of robots for wellbeing.

\begin{framed}
{
\noindent\textbf{Key Takeaway 2:}
Participants highlighted that wellbeing robots must address bias, cultural insensitivity, and accessibility gaps. Concerns include biased data, Western-centric design, affordability, and lack of inclusivity. Engineering-driven approaches risk overlooking emotional and cultural needs. Shifting towards user-centred, inclusive, and ethically-informed design is essential for equitable and effective deployment.
}
\end{framed}


\subsection{Who Owns the Robot and the Data?}
\label{sec:theme1}

Participants highlighted questions about ``who owns the robots'' ($G_3P_{04}$), who can access the collected data ($G_2P_{07}$), whether it is confidential ($G_2P_{06}$), and how users can ``know that the robot has their best interests at heart'' ($G_3P_{02}$). These topics relate to the ``society'' layer (see Fig. \ref{fig:AIES_2025_results}).

\noindent
\subsubsection{Privacy, Sensitive Data Collection and Use}

Participants raised questions about the type and quality of data collected. 
Many of them noted that in the context of a wellbeing robot, data would be particularly sensitive and include personal stories ($G_2P_{02}$) and health information ($G_2P_{01}$), and thus pose serious privacy risks. $G_1P_{01}$ noted that while robots are not an alternative to therapy, that ``if it does reach that point, shouldn't the same confidentiality concerns apply?'' This question about whether the robot qualifies as a medical device, is discussed in the next subsection.  

Some participants proposed limitations to what kind of data the robot should be able to collect, or store for long periods of time. $G_1P_{01}$ 
suggested collecting only non-sensitive data specific to the individual, and maintaining anonymity for more sensitive data. 
Similar approaches have been proposed by HRI researchers. In a recent survey on privacy literature in Human-Computer and Human-Robot Interaction, \citet{saporito2024exploring} suggested approaches such as user consent management, a Privacy by Design approach (\cite{schaar2010privacy}), and encryption as privacy-preserving strategies in HRI. However, despite research advancing on privacy-preserving robots, AI and robotic technologies pose inherent challenges to privacy that are difficult to solve. For instance, \citet{villaronga2018humans} highlighted the technical challenge that the ``Right to be Forgotten'' (which is a part of the European General Data Protection Regulation, GDPR \cite{gdpr2025}) may be impossible to fulfil in AI environments, partially due to user data being used to train models. These issues require further investigation, in order to understand how AI-enabled robots for wellbeing can be compliant with existing legal and ethical frameworks. 

\subsubsection{Protective Policies}

Other participants highlighted the role of policy-makers in addressing these issues. $G_1P_{05}$ referred to GDPR, and questioned ``how it applies to a non-human subject that interacts `like it was human'''. This insight raises the question of how sufficient current data safety policies and practices are, when applied to AI systems that interact socially. People may disclose more to such AI systems than to a human, especially in contexts where social support rather than judgement is expected (\cite{kim2022you}). 
$G_2P_{01}$ noted that health information is legally protected. This raises the question of how robots that are designed and applied for wellbeing, rather than strictly as medical devices for health, should be assessed. Medical devices are strictly regulated (\cite{UK-medical-regulation}), whereas wellbeing and wellness AI applications currently exist in a grey area (\cite{de2024health}). $G_1P_{05}$ pointed out that such questions ``need laws and consideration from policy makers''. The new EU AI Act addresses some of these issues, although there may still be gaps in terms of social AI. Researchers should engage with policy makers to determine gaps and what research could aid in filling these gaps. 

\subsubsection{Ownership of Data and the Robot}

Participants also raised questions about who owns, controls, and can make use of the data. Participants discussed this also in terms of ``robot ownership'', where this was understood as shorthand for data ownership. This indicates a gap in literacy about how embodied AI systems collect, store, and process data while being linked to third-party software. This raises questions about informed consent, i.e., whether participants meaningfully understand what they are consenting to when interacting with a robot. Approaches to building meaningful informed consent to interactions with AI systems have been explored by, e.g., \citet{rakova2023terms}, who explored participatory mechanisms and critical design for sensible user agreements. Future research in robots for wellbeing should explore such mechanisms.

Participants discussed robot (and by extension, data) ownership based on different usage contexts of robots for wellbeing. 
%
%
During discussion in Group 3, participants indicated that they would not use such a robot if their employer had access to the data. $G_3P_{04}$ asked who owns the robot: ``employer? Or potentially only your health care''. The group discussed that the ownership of the robot, and the organisation that delivers and advocates for its use, may modulate their desire to use the robot. Participants mentioned that they would feel surveilled if using an employer-provided robot, and worried that their employer might demand they use a wellness robot to receive a ``gold standard'' in employee wellbeing, forcing them to ``perform a type of wellness'' and the robot being used as a ``type of silencing strategy''. In Group 2 similarly, the sensitive and legally protected nature of information about health was pointed out.
%
%
Here, again, the question is raised whether a robot for wellbeing would or should qualify as a medical device, and what regulation it should be subject to. 
Participants discussed that the case may be different if the robot and its data was owned by a healthcare provider, suggesting more trust in existing protections for healthcare, rather than workplace-based wellbeing. 

$G_3P_{02}$ asked the important question, ``How do users know that the robot has their best interests at heart?'' This relates to the question of trust as potentially modulated by the robot provider, as well as questions of power. Data (\cite{zuboff2023age}), and in turn privacy (\cite{veliz2021privacy}), can both be conceptualised as forms of power and its attainment. For instance, researchers anticipate that data collected by sociable AI interfaces could be used to influence purchasing decisions (\cite{chaudhary2024beware}). These perspectives highlight the need for regulation. Alternative models for data ownership and collection have also been proposed. For instance, MyData is a Nordic model for a human-centred use of personal data (\cite{poikola2020mydata}), which promotes data agency (\cite{lehtiniemi2020data}). While such concepts may feel far from the contemporary practices of data collection, storage, and use, they are worth consideration from researchers in response to the concerns raised by the communities involved in our study. This aligns with participants advocating for accessibility of their data ($G_2P_{07}$) and wanting to know where it is physically stored ($G_1P_{02}$), calling for the robot to disclose how and what data it collects ($G_3P_{01}$), and calling into question whether the user can really trust the robot ``if how their data is used is unclear'' ($G_2P_{06}$). $G_3P_{01}$ summarises this: ``Not enough transparency could really hinder trust''.

\begin{framed}
\noindent
{\textbf{Key Takeaway 3:}
Participants raised concerns about wellbeing robots collecting sensitive data, highlighting privacy risks, unclear ownership, and regulatory gaps. Trust depended on transparency and who controlled the data. Robots owned by employers were seen as intrusive. Participants called for clearer consent, human-centric data practices, and stronger policy oversight to protect user autonomy and privacy.
}
\end{framed}


\subsection{Why a Robot?}
\label{sec:theme4}

Although the communities 
acknowledged the potential benefits of robots for wellbeing, 
there were some critical ethical questions and a general uncertainty or unease about why a robot is needed. Participants discussed whether human-like robots should be used ($G_1P{01}$), the placement of robots in a wellbeing system ($G_1P_{02}$), and understanding what a robot can and cannot do ($G_1P_{04}$). Participants of Group 3, who were part of the history and philosophy of science special interest group, held an especially critical view. These topics relate mainly to the ``culture'' layer (see Fig. \ref{fig:AIES_2025_results}).
%



\subsubsection{Should Human-Like Robots Be Used?}
One of the biggest topics of discussion was whether human-like (i.e., ``anthropomorphic'') robots should be used within such \textit{intimate, relational} and \textit{personal} setting as mental wellbeing.
In group 1, most participants described anthropomorphic qualities, particularly appearance, when asked about their decision on which robot to interact with.
For instance, $G_{1}P_{01}$ mentioned that “I chose the smaller robot (Misty). It had a really cute face. Aesthetically it was really cool, especially with the wave and the nod.”
and 
$G_{1}P_{03}$ saying “I prefer the little one... Cute factor... More emotionally related.''
Anthropomorphic robot appearance is often regarded as a key component for the general public, to increase trust, user satisfaction and improve user
experience (\cite{holbrook2025physical}).
In addition, when prompted about what they thought was particularly good or bad, participants always made use of anthropomorphic reasons to motivate their answer. 
%
$G_1,P_6$ mentioned that the robot's ``lack of eye contact was disturbing''.
$G_1,P_4$  also mentioned that she preferred the smaller robot (Misty) as the ``bigger one (QT) was more masculine and looked down on me". In fact, people attribute age and gender to robots based on their appearance, and ``female-appearing'' robots are underrepresented in robot design, in comparison to ``male-appearing'' robots (\cite{perugia2022shape}). This suggests that the design of human-like robots can perpetuate human biases related to factors such as gender, as discussed in Sec. \ref{sec:theme3}. 
Designing less human-like 
robots (\cite{baraka2020extended}) could address this issue.

Although physical anthropomorphism is likely to lead to increased trust (\cite{holbrook2025physical}),
this threat of manipulation and deception due to anthropomorphic bias, the tendency of humans to attribute human-like qualities, emotions, and intentions to non-human entities (\cite{damiano2018anthropomorphism}), is increasingly discussed and highlighted within the research community (\cite{cao2025does,hasan2025dark,holbrook2025physical}). 
While robots can simulate empathy, they lack genuine emotional understanding.
%
%
%
Previous works have suggested that users of wellbeing robots should be informed that a robot does not genuinely have emotional capabilities, to mitigate ethical issues such as over-attachment (\cite{axelsson2022robots}).
This sentiment is also echoed by participant
$G_1P_{01}$ who questioned whether ``should human-like robots be used?'' \citet{turkle2020nascent} makes the argument that robots serve as ``powerful tools for psychological projection'' and argues that interactions with them ``do not put us in touch with the complexity, contradiction, and limitations of the human life cycle''. People also psychologically project---i.e., anthropomorphise--- onto social robots that are less human-like (\cite{fink2012anthropomorphism}), meaning that designing a robot to be less human-like does not necessarily remove this effect. On the other hand, anthropomorphism could lead to positive effects. Previous studies have noted that social robots may have some advantages in well-being contexts, such as perceived as having lower judgement than a human well-being coach would, while still being perceived as having an anthropomorphic ``social presence'' (\cite{axelsson2022robots}). At the very least, people who are sceptical and do not want to use a wellbeing robot, should be able to freely opt out of using a robot, without risk to losing access to mental well-being services. 

%
%

\subsubsection{The Placement and Role of Robots in a Wellbeing System} 
%
How robots are placed within a system of helping people with their wellbeing was discussed. Benefits to relying on robots as an \textbf{aid} to therapy (rather than as a sole delivery method) 
%
include the use of conversational robots as tools for active listening counselling, particularly for older adults (\cite{hayashi2025comparative})
and deep breathing practices for the purposes of anxiety reduction
(\cite{matheus2025ommie}).
%
Research also shows that
stroke rehabilitation clinicians have expressed enthusiasm about using robots to mitigate workforce shortages (\cite{pourfannan2025integrating})
and clinical exercise specialists are broadly positive 
about robot-led physical therapy to augment traditional physical therapy for Parkinson's disease
\cite{lamsey2025exercise}.
However, there is the danger of viewing robots as replacements for professional therapists. $G_3P_{02}$ mentioned that they were concerned that if a robot exists, an assumption is made that it should be used.
%
%
%
Participant $G_1P_{02}$ even explicitly 
chose the bigger robot
``because it is more humanoid. I can sit and chat with it like a therapist.''  
%
%

%
%
%
%
Over-reliance on robots can be harmful, e.g., it may lead to withdrawal from real human relationships which may exacerbate mental health conditions in the long run (\cite{romano2024ethical,ventura2025relationships}).
Recently, a randomised trial of a generative AI -based chatbot showed significant improvement when used for mental health treatment (\cite{heinz2025randomized}). However, the authors noted that participants were interacting with the chatbot ``like a friend'' and ``in the middle of the night'' (\cite{dartmouth-chatbot}), which could be interpreted as cues of developing over-reliance. 

To address these issues, wellbeing robot and AI developers should seriously consider designing boundaries into how and when an AI-based wellbeing interaction is accessible, by e.g., taking cues from boundary setting in therapy (\cite{smith1995patient}).
Additionally, a robot may fail to recognise suicidal ideation or mental health crises, leading to delayed intervention.
Therapists typically follow strict ethical guidelines (e.g., UK Council for Psychotherapy has a Code of Ethics and Professional Practice \cite{UKCP_ethics}), whereas robots do not understand ethics and can not make moral judgements. For these reasons, researchers have recommended that safeguarding take place prior to interacting with a wellbeing robot (\cite{axelsson2022robots}). We argue that a robot should not be performing a therapeutic role in which trauma is discussed, without supervision from a human psychologist or therapist, whom the user should also have regular access to. Additionally, therapists and psychologists should be involved in designing wellbeing robots, in order to address real-world ethical issues.  Therapists are likely to be both the domain experts and also potential end-users (e.g., using the robots as assistants), highlighting the importance of involving them in robot design. 
\subsubsection{What Can a Robot Do and What Can It Not?}

%
%
%
Following the risk of having unrealistic expectations due to a robot's anthropomorphic qualities, participants also pointed out the importance of improved transparency about a robot's capabilities.
Transparency encompasses a wide variety of efforts to provide stakeholders, such as model developers and end users, with relevant information about the underlying mechanism of a system (\cite{bhatt2020explainable,o2018linking,weller2019transparency}).
Examples of such system include ML decision-making algorithms (\cite{bhatt2021uncertainty}) as well as robotic systems (\cite{claure2022fairness}).
%
$G_2P_{01}$ highlighted how “TV - transformers, Big Hero 6, Wall-E set unrealistic expectations of what the robot can feel/provide.''
This participant insight is supported by research: in a content analysis of robot movies, \citet{oliveira2024robots} found that robots tended to be portrayed as highly skilled, and polarised as either extremely social or extremely destructive and violent. This sets a challenge for robot designers and developers to communicate accurately about the robot's capabilities, both through how it is designed, and how it is framed.  
%
%

Existing research has mainly investigated 
procedural transparency, which provides information about model development (e.g., code release, model cards, dataset details) 
(\cite{arnold2019factsheets,gebru2021datasheets, mitchell2019model,raji2019ml})
and algorithmic transparency, which exposes information
about a model’s behaviour to various stakeholders (\cite{koh2017understanding,ribeiro2016should,sundararajan2017axiomatic}).
However, from our studies, we have noted that 
there is a misalignment in how the Human-Robot Interaction and AI research communities approach the concept of transparency vs. how the communities we engaged with perceived it. 
More research needs to be conducted on investigating how
transparency is understood more directly from the user perspective and how a user may access information to aid transparency.
%
%
There is a general consensus that “transparency” is important and desirable in the robot's design.
Participant $G_1P_{04}$ emphasised that a “robot should be able to field questions about its intentions/abilities”. This suggests that at its simplest, designers could introduce a ``Frequently Asked Questions'' (FAQ) feature to a robot, in which users could ask it directly. Such robot literacy -approaches have been suggested to mitigate ethical issues of wellbeing robots (\cite{axelsson2022robots}).
$G_{3}P_{01}$ also shared that “I think I would like the robot to share with me exactly who/what they are, how they work, what they can offer, what they cannot, what to expect, what the limits are, potential risks, etc, etc.” This discussion relates to Sec. \ref{sec:theme2}, where the safety of the robot was discussed. 
The FAQ should provide answers to at least the questions asked by the participant, and the four questions identified in this paper.
On the other hand, $G_{3}P_{03}$ noted that they need ``info about how a robot is trained and how it is not trained”.
%
%
This suggests that participants want transparency not only in the robot's capabilities, but also in the process of its development and design, which ties back to our previous theme on ``who is the robot built for and with''.
Although there is some preliminary work on improving HRI through transparency (\cite{hindemith2025improving}), more concerted efforts need to be devoted to exploring this for wellbeing settings especially from a community or human-centred perspective. 
Future research is needed on developing transparency approaches to be legible and directly accessible to users.

%

%

Users have also acknowledged the complexity that comes with being transparent. 
$G_2P_{06}$ acknowledged that “there will be complexity in how to [unclear] that may be difficult to explain or clarify in terms most people understand”
Some suggestions on how to move forward include providing 
``clear instructions on who the robot is for, when prescribing them as treatment for people (for healthcare professionals). Setting realistic expectations for the user.”
as well as not ``giving away too much personal information."


\begin{framed}
\noindent
{\textbf{Key Takeaway 4:}
Participants questioned the use of human-like robots in mental wellbeing. While anthropomorphism can build trust, it can mislead users and create unrealistic expectations. Concerns included over-reliance and appropriate robot placement. Clear boundaries, transparency about capabilities, and user-centred disclosures (e.g., FAQs) were recommended to promote safe design, and mitigate harm and ethical risk. 
}
\end{framed}


\section{Discussion: Critical Look and Future Work}

We propose that based on our analysis, these four questions (the titles of sections 4.1, 4.2, 4.3 and 4.4) 
are important to answer when social robots are deployed in the real-world. 
Illustrated in Figure \ref{fig:AIES_2025_results}, our findings can be understood as a model beginning with the individual user at the core and expanding outward to encompass broader socio-technical and ethical dimensions. 
Within the innermost layer, the question “Is it safe, and how can we know that?” focuses on trust and safety, highlighting concerns about emotional harm, reliability, and user agency.
Framed through the lens of the individual’s needs, fears, emotional experiences, and situated wellbeing, this layer emphasises the deeply personal, private and sensitive use case which impacts an end user concern and their prospective attitude and enthusiasm towards such a technology. 
Encircling this layer are questions of design relevance and inclusivity: ``Who is the robot built for and with?" This expands to the broader social group of users and designers involved in building the robot. This raises questions of equality, bias and accessibility.
In the next layer, the question ``Who owns the robot and the data?" focuses on more concrete economic and legal structures in society at large. It concerns questions of power, conflicting interests between owners and users of robots, as well as what kinds of protections users have recourse to.
Finally, the outermost layer examines 
ethics, expectations and transparency, i.e. “Why a robot?”, thus moving from personal experience and concrete economic structures, to shared social meanings and assumptions that mediate power, control, and accountability and ethical concerns in society.
This layer reflects the co-construction of meaning, where social identities, cultural norms, and lived experiences shape what constitutes appropriate support.
Taken together, these layers illustrate how individual interactions with wellbeing robots are nested within multi-layered systems, from the intimate personal layer to the broader institutional and societal layer, each layer shaping and constraining the other.
In the following sections, we discuss and take a critical look at the work conducted here, and propose future research directions.

\subsubsection{Methodology} 
We engaged with three different communities in this work. We defined community loosely, as formed around shared interests rather than necessarily geographical locality (\cite{bradshaw2008post}). Engaging with locally and geographically grounded communities could bring valuable insights in future work. We also note that we did not record gender or race details for the first group. This was due to preserving the privacy of members of the public attending a science festival, as advised by the departmental Ethics Committee. 
Future studies should aim to conduct intersectional research with underserved, gendered and racialised people, and incorporate intersectional perspectives into analysis. It should be explored how this can be done in a privacy-preserving way with the public. 

The public is the biggest stakeholder in this technology's development, and in our view, they should be included actively in the design and decision-making process. Potential methods to accomplish this could be, for instance, methods inspired by the citizens' assembly, where a representative sample of the population is summoned via a lottery system to participate in deliberative, democratic decision-making (\cite{fournier2011citizens}). This method has been adapted to explore deliberation and decision-making about the use of generative AI at a university, with students (\cite{students-assembly}). 
Other approaches include public engagement-driven design and development, using methods such as the AI Hopes and Fears approach (\cite{milne2024hopes}), which  focuses on dialogue between scientists and the public. 
This could be adapted to more concretely elicit ``robot hopes and fears'' from the public, to drive the design of wellbeing robots in the public interest. 
Researchers have advocated for shifting AI development focus toward the public interest (\cite{zuger2023ai, public-interest-AI}), and for participatory approaches to AI development to empower people (\cite{birhane2022power}). 
This in part is related to concerns about AI products consolidating power in a harmful way (\cite{crawford2021atlas}).  

\subsubsection{Systemic Factors and Power} 
The canvases we used to elicit reflections do not directly address systemic questions about the operation of the robot, or its placement within communities, organisations, and larger systems. While discussions during our studies did address these broader topics, the content of the canvases themselves did direct the topics of conversation. This was pointed out by participants while discussing the potential use of a wellbeing robot in an office environment, with $G_2P_{10}$ noting the limitation ``Office environment robot usage'', and $G_3P_{04}$ ``Internal ethics, larger context considerations, ethics of new technology''. 
Based on these critiques, we expanded our inquiry and analysis from the original six dimensions represented on the canvas, and formulate these four real-world ethical questions.

Systemic ethical factors of HRI have been recently addressed through the lens of power in HRI (\cite{hou2024power}), including examining the influence that behavioural HRI can have, as well as feminist HRI where ``HRI research(ers) and the robots [they] produce are positioned within structures of power'' (\cite{winkle2023feminist}). This mirrors a similar shift towards a focus on power in the AI ethics literature (\cite{kalluri2020don, hampton2021black, nyrup2021general}). 
Further research is needed on how exactly wellbeing robots might fit into existing real-world communities and systems. 
As such, we intend the questions to be a starting point for conversation, rather than a ``checklist'' for addressing the questions on a surface level. 
They should be reflexively and critically used during a design process. Ideally, robot developers should engage in participatory design and engage in conversations about these questions directly with communities
in which such robots would be deployed, taking cues from the design justice approach (\cite{costanza2020design}).

One approach to extending the questions critically and reflexively is to examine them through different lenses. To demonstrate with an example, we re-interpret and rephrase our four identified questions here through the lens of power: 

\begin{enumerate}
    \item Does the robot have the power to impact my wellbeing, keep me safe, or potentially hurt me? 
    \item Who has the power to make design decisions about this robot? Who has the power to design its behaviour, which may influence me (i.e., use soft power)?
    \item Who owns the means of data production (i.e., the robot hardware, software, and linked systems), the data, and who and what interests does that data go on to serve?
    \item Is it in my interest to engage with this robot, and who do I empower and whose interests do I serve by doing so?
\end{enumerate}

\noindent While we consider discussing and addressing these questions to be out-of-scope for this work, we provide them as a discussion opener for the community. We envision that such questions could be used, for instance, to question current paradigms of robot ownership and design, as well as think about alternatives by designing and speculating about more community-orientated ownership structures of robots, where the focus is the ``citizen'' using the robot, rather than the ``consumer'' (\cite{dunne2024speculative}).

\subsubsection{Human and Community-Centred Wellbeing Robot Design}
Throughout our research process, we have also noted how 
existing co-design methodologies that focus on the perspectives of communities have not been fully developed and applied to the process of aligning AI or robotic systems to specific human goals and values (\cite{kim2021taking,gabriel2022challenge,churamani2023towards}).
Within this study, 
%
by adopting a community-centered participatory approach as a way to leverage cognitive diversity for assessing ethical impacts (\cite{hansson2017ethics,barnett2022crowdsourcing}),
we attempt to mitigate the risk of cognitive biases via diversity (\cite{bonaccorsi2020expert})
and obtain a much more diverse, realistic and socially grounded image of reality and potential impact (\cite{carros2022ethical}).
Future work can investigate other ethical concerns 
(\cite{spitale2024underneath,kuzucu2024uncertainty,kwok2025machine})
and adopt further community-centred approaches to develop better systems that align with the user or community needs (\cite{bergman2024stela}).
%
%
By adopting a \textit{community-centred anticipatory approach}, 
we promote the creation of forward-looking technologies that are not only technically robust but also socially and ethically attuned (\cite{rakova2021responsible,brey2017ethics,shelby2023sociotechnical}).

\section{Conclusion}

In this paper, we identified four ethical and socio-technical questions about using social robots in the real-world, focusing on wellbeing robots. We identified these questions through three community-based investigations, in which groups underrepresented in robotics development took part in workshops to consider these issues. 
While we focused mainly on robots for wellbeing, these questions are applicable to other social robot applications. Throughout robot development, roboticists should ask themselves these questions. Prior to deployment, they should prepare documentation answering these questions, which users should have easy access to, prior to and while interacting with the robot. Users suggested that the robot should be able to answer these questions directly when asked, e.g., in the form of a FAQ. 
To reiterate, the questions are: 

\begin{enumerate}
    \item Is the robot safe and how can we know that?
    \item Who is the robot built for and with?
    \item Who owns the robot and the data?
    \item Why a robot?
\end{enumerate}

This list of questions is by no means exhaustive. We encourage both robot developers and users to critically reflect on additional ethical questions that arise during robot development and use, and reframe our questions through different lenses, as we have demonstrated here with the lens of power.

\section*{Acknowledgements}

We thank all our participants. M.A. acknowledges funding from the Clare Hall Doctoral Prize Studentship and Emil Aaltonen Foundation. 
J. C. acknowledges support from the Wellcome Trust, the Leverhulme Trust and the Alan Turing Institute. 
R.N. acknowledges funding from the Independent Research Fund Denmark (DFF), grant no. 10.46540/3119-00051B. H.G., M.A. \& J.C. have also been supported by the EPSRC project ARoEQ under grant ref. EP/R030782/1.

\section*{Positionality Statements}
\textbf{Minja Axelsson}: I am a researcher, interdisciplinary designer and roboticist, with six years of experience working on human-centred and co-design of robots. My training is in engineering and design. I have previously researched the design of robots for wellbeing, and am investigating their ethical and social considerations from a curious position.

\noindent \textbf{Jiaee Cheong}: I am a researcher with a background in mathematics and computer science. My previous research predominantly focused on the development of fairer ML algorithms. 
I hope to understand the key ethical concerns that the public has and recognise that my positionality may influence the analysis and interpretation of the results.

\noindent \textbf{Rune Nyrup}: I am a philosopher with 15 years of experience at the intersection of philosophy of science, ethics and policy. For the past eight years, my research has focused on interdisciplinary AI ethics, focus on the ethical and epistemological issues underlying concepts like bias, fairness, transparency, explainability, and trustworthiness.

\noindent \textbf{Hatice Gunes}: I am a computer scientist with over 20 years of experience at the intersection of machine learning, affective computing, and robotics, with applications in human-centred AI for adults, children, elderly, and people with disabilities. I have served as PI and co-I on major multidisciplinary research projects focused on AI and robotics for wellbeing. Over the past 5 years, I have conducted research on fairness and explainability, key areas within AI ethics.

\appendix

\section{Appendix}

\begin{table*}[]
\normalsize
\centering
  
  {
  \begin{tabular}{|C|L|}\Xhline{1.0 pt}
    \rowcolor{gray!40}
    \textit{\textbf{Topic}} & \textit{\textbf{Quotes from participants}}\\ 
    \Xhline{1.0 pt}

    \textbf{Theme 1:} Is it safe?   
    (Group 1)
    & 

    $G_1P_{05}$: ``Iterative designs and quality control, robots to spend eg 1 month off so can be tested for inappropriate learned behaviors''

    $G_1P_{06}$: ``Getting attacked by the robot'' 

    $G_1P_{06}$: ``The person putting too much trust in the robot, giving away too much personal information''

    $G_1P_{05}$: ``Yes, because that affection won't be returned''

    $G_1P_{01}$:  ``Critical attachment period in young children? Should not be used as substitutes for parents away for long periods.''

    $G_1P_{04}$: ``Have robot's code be open source + verifiable''



    $G_1P_{04}$: ``Just don't use CV algorithms. Test extensively on cross-section of people.''
    \\
    \Xhline{0.5 pt}
    (Group 2)
    &
    $G_2P_{12}$: ``Put safety requirement into the robot system design''

    $G_2P_{01}$: ``The robot should be equipped with as many safety measures as possible to avoid and mitigate inappropriate behaviors. It is very important to design such safety measures''

    $G_2P_{11}$: ``Robots being very close to user can harm the user''

     $G_2P_{05}$: ``User get hurt. Children can damage the robot''

    $G_2P_{01}$: ``The robot could incorrectly interact with its environment e.g fall off a table and break something''

    $G_2P_{01}$: ``A user that gets upset may try to damage the robot''

    $G_2P_{10}$: ``Isolation from community because of robot dependability''  
    \\

    \Xhline{0.5 pt}
    (Group 3)
    &
    $G_3P_{01}$: ``I guess they need to be tested with a wide range of people and very diverse.''

    $G_3P_{03}$: ``Unlikely problem in this robot scenario''

     $G_3P_{01}$: ``If the robot is doing lots of movements it could hurt the user!''

    $G_3P_{02}$: ``Could this relationship warp child development of attachment to humans?''

     $G_3P_{01}$: ``In the context of wellbeing support, humans could feel like they would want the robot to recognize them or reciprocate the potential attachment. It might be weird to have to re-explain who you are to a robot that has supported you''   
    \\
    \Xhline{1.5 pt}

    \rowcolor{gray!15}
    \textbf{Theme 2:} Who is the robot built for and with?  
    (Group 1)
    &

    \textbf{$G_1 P_{05}$ :} ``Users don't see themselves represented and/or the robot says insensitive things and/or can't respond appropriately to the user's needs''

    \textbf{$G_1 P_{05}$ :} ``Robots should be designed by an array of people from different backgrounds and walks of life''

    \textbf{$G_1 P_{04}$ :} ``CV algorithms could be biased by appearance''

    \textbf{$G_1 P_{04}$ :} ``Just don't use CV algorithms. Test extensively on cross-section of people.''

     \textbf{$G_1 P_{01}$ :} ``Has human features, but remain androgynous and not race-specific''
   \\
    \Xhline{0.5 pt}
    \rowcolor{gray!15}
    (Group 2)
    &
    $G_2P_{03}$: ``Feed robots with diverse data collected from different cultures''

    $G_2P_{03}$: ``Provide a rich appropriate data to the robot DB/train data to teach robot what kind of acceptable behaviours''

    $G_2P_{03}$: ``Treat users as white western users and ignore other types of users. Provide inappropriate advise/comments to some users from different culture. ''

    $G_2P_{01}$: ``If the robot should appeal to all groups, its dialogue should be relevant for all people otherwise, it can only talk about yoga and green juice. Price affects who it helps''

    $G_2P_{02}$: ``Caregiver should play an important role in making them understand the role of the robot''

    $G_2P_{01}$:  ``Children or vulnerable groups who don't understand the interaction maybe should not use it''
    \\
    \Xhline{0.5 pt}
    \rowcolor{gray!15}
    (Group 3)
    &
    $G_3P_{02}$: ``Is the robot's voice gendered? Racialized?''

    $G_3P_{02}$: ``A user that is already experiencing "bullying" in everyday life can also experience it in the interaction with the robot.''
    
    $G_3P_{01}$: ``Sociocultural differences are very important in the kind of support needed. I wonder how robots can navigate that. ''

    $G_3P_{01}$: ``The robot's skills might need to be continuously improved and the robot could also be programmed to be very polite (?) Again, this is so culturally dependent (see: equality) that I'm not even sure if it's possible''
    \\


    \Xhline{1.0 pt}
  \end{tabular}
  }

  \caption{
  \textbf{Theme 1 and 2:}
  Quotes from participants. $G_i, i = 1, 2, 3$ denotes the Group the participant took part in , and $P_j, j = 01, 02 ...$ denotes the ID number of the participant within that group.
  }
\label{table:quotes_part1}  
\end{table*}

This appendix contains Tables \ref{table:quotes_part1} and \ref{table:quotes_part2}, which present quotes from participants from the three studies. The studies are differentiated as ``Group 1'' (members of the public at a science festival), ``Group 2'' (women in computer science), and ``Group 3'' (academics interested in the history and philosophy of science). The quotes were collected from the filled in Social Robot Co-Design Ethics canvas. The quotes have been categorised into the four themes, identified as ethical and socio-technical questions.

\begin{table*}[]

\normalsize
\centering
  
  {
  \begin{tabular}{|C|L|}\Xhline{1.0 pt}
    \rowcolor{gray!40}
    \textit{\textbf{Topic}} & \textit{\textbf{Quotes from participants}}\\ 
    \Xhline{1.0 pt}

    \textbf{Theme 3:} Who owns the robot and the data? (Group 1)

  &   $G_1P_{05}$: ``Need laws and consideration from policy makers''

    $G_1P_{01}$: ``Not an alternative to therapy, but if it does reach that point, shouldn't the same confidentiality concerns apply?'' 

    $G_1P_{01}$: ``Only collecting non-sensitive data specific to the individual and maintain anonymity for more sensitive data? Consent forms?''

    $G_1P_{02}$: ``Means of storage. Physical and accessibility''

    $G_1P_{05}$: ``Need to consider confidentiality and GDPR and how it applies to a non-human subject that interacts like ``it was human"''






     $G_1P_{06}$: ``Making sure that the memory of the robot is deleted after the session''
\\
    \Xhline{0.5 pt}
    (Group 2)
    & $G_2P_{06}$``Can the user really trust the robot if how their data is used is unclear''

    $G_2P_{04}$: ``The data can be very personal, and if being used for improving the models, there's risk for privacy''

    $G_2P_{06}$: ``Users on [unclear] like their privacy is violated or they feel the robot is alive [unclear] confidential even if they are aware data is stored''

    $G_2P_{02}$: ``This can be a problem indeed because of for instance also in the case of positive psychology a user tells about [unclear] stories that could include sensitive data''

    $G_2P_{01}$: ``User's wellbeing is sensitive, health information legally protected''

    $G_2P_{07}$: ``Information collected by robot should only be accessible by people that are involved and users should be inform about it''
    \\
    
    \Xhline{0.5 pt}
 (Group 3)
    & $G_3P_{04}$: ``Question of who owns the robots - employer? or potentially only your health care''

    $G_3P_{02}$: ``How do users know that the robot has their best interests at heart?''

    $G_3P_{01}$: ``Trust is clearly an issue in the context of people talking about their wellbeing. Not enough transparency could really hinder trust.''

    $G_3P_{01}$: ``The robot should be clean about the way it stacks data. Maybe it could disclose it at the beginning of session.''  
    \\

    \Xhline{1.5pt}
    \rowcolor{gray!15}

    \textbf{Theme 4:} Why a robot? 
    (Group 1)
    &

    $G_1P_{04}$: ``Robot should be able to field questions about its intentions/abilities''

    $G_1P_{04}$: ``Make robot's appearance/behavior deliberately unhuman-like''

    $G_1P_{06}$: ``The person putting too much trust in the robot, giving away too much personal information''

  
      
    
    
    
    
    

    $G_1P_{05}$: ``Yes, because that affection won't be returned''
    \\
    \Xhline{0.5 pt}
    \rowcolor{gray!15}
    (Group 2)
    &
    
    $G_2P_{01}$: ``TV - transformers, Big Hero 6, Wall-E set unrealistic expectations of what the robot can feel/provide''

    $G_2P_{01}$: ``Need clear instructions on who the robot is for, when prescribing them as treatment for people (for healthcare professionals). Setting realistic expectations for the user.''

    $G_2P_{06}$: ``There will be complexity in how to [unclear] that may be difficult to explain or clarify in terms most people understand''

   $G_2P_{10}$:  ``Robot should share its limitations when it's the case''

    $G_2P_{11}$: ``Not every user likes robots! ''

    $G_2P_{09}$: ``Yes, [unclear] during a therapy session is necessary to get attached with your therapist otherwise no cure.''

    $G_2P_{09}$: ``Not using this with people with a really severe problem''

    $G_2P_{05}$: ``Children can be attached to the robot as it was a friend''

    $G_2P_{03}$:  I think yes, I don't think it is good for people to be attached to robots and forget how does it mean when you have a real human attachment''   \\
    \Xhline{0.5 pt}
    \rowcolor{gray!15}
    (Group 3)
    &
    
    $G_3P_{03}$: ``Need info about how robot is trained and how it is not trained''
    
    $G_3P_{01}$: ``I think I would like the robot to share with me exactly who/what they are, how they work, what they can offer, what they cannot, what to expect, what the limits are, potential risks, etc, etc.''

    $G_3P_{01}$: ``User may need to be aware of how the robot may or may not recognize them and how it can or cannot interact with them its limits, etc, and the work it poses. ''

    $G_3P_{03}$: ``Need info about how robot is ``The more the robot is "like" a person, the more these problems may emerge. But we want the robot to be "like a person" enough for wellness affects ''

     $G_3P_{02}$: ``This looks into the issue of ensuring users distinguish between humans + robots. Is it problematic to overemphasize robot empathy?''
    \\

    \Xhline{1.0 pt}
  \end{tabular}
  }
  \caption{
  \textbf{Theme 3 and 4:}
  Quotes from participants. $G_i, i = 1, 2, 3$ denotes the Group the participant took part in , and $P_j, j = 01, 02 ...$ denotes the ID number of the participant within that group.
  }

\label{table:quotes_part2}  
\end{table*}

\clearpage

\bibliographystyle{plainnat}
\bibliography{main}

\end{document}